# Stable Acceleration of a LHe-Free Nb₃Sn demo SRF e-linac Based on Conduction Cooling

**Ziqin Yang**[a,b,c], **Yuan He**[a,b,c*], **Tiancai Jiang**[a], **Feng Bai**[a], **Fengfeng Wang**[a], **Weilong Chen**[a], **Guangze Jiang**[a,b], **Yimeng Chu**[a,b], **Hangxu Li**[a,b], **Bo Zhao**[a], **Guozhen Sun**[a], **Zongheng Xue**[a], **Yugang Zhao**[a], **Zheng Gao**[a], **Yaguang Li**[a], **Pingran Xiong**[a], **Hao Guo**[a], **Liepeng Sun**[a], **Guirong Huang**[a,b], **Zhijun Wang**[a,b,c], **Junhui Zhang**[a,b,c], **Teng Tan**[a,b,c], **Hongwei Zhao**[a,b], **Wenlong Zhan**[a,b]

[a]Institute of Modern Physics, Chinese Academy of Sciences, Lanzhou 730000, China

[b]School of Nuclear Science and Technology, University of Chinese Academy of Sciences, Beijing 100049, China

[c]Advanced Energy Science and Technology Guangdong Laboratory, Huizhou 516007, China

* Corresponding author at: Institute of Modern Physics, Chinese Academy of Sciences, Lanzhou 730000, China

E-mail address: hey@impcas.ac.cn

**Abstract**

The design, construction, and commissioning of a conduction-cooled Nb₃Sn demonstration superconducting radio frequency (SRF) electron accelerator at the Institute of Modern Physics of the Chinese Academy of Sciences (IMP, CAS) will be presented. In the context of engineering application planning for Nb₃Sn thin-film SRF cavities within the CiADS project, a 650MHz 5-cell elliptical cavity was coated using the vapor diffusion method for electron beam acceleration. Through high-precision collaborative control of 10 GM cryocooler, slow cooldown of the cavity crossing 18K is achieved accompanied by obviously characteristic magnetic flux expulsion. The horizontal test results of the liquid helium-free (LHe-free) cryomodule show that the cavity can operate steadily at $E_{pk}$=6.02MV/m in continuous wave (CW) mode, and at $E_{pk}$=14.90MV/m in 40% duty cycle pulse mode. The beam acceleration experiment indicates that the maximum average current of the electron beam in the macropulse after acceleration exceeds 200mA, with a maximum energy gain of 4.6MeV. The results provide a principle validation for the engineering application of Nb₃Sn thin-film SRF cavities, highlighting the promising industrial application prospects of a small-scale compact Nb₃Sn SRF accelerator driven by commercial cryocoolers.

## 1. Introduction

In 1911, Onnes discovered that the DC resistance of Hg suddenly drops to zero at 4.2K, marking the beginning of superconductivity research. For over a century, the pursuit of superconducting materials with higher transition temperatures has been the focus of basic research in superconductivity physics. Utilizing superconducting materials with higher transition temperatures in engineering technology represents the direction of applied basic research. The application of high-temperature superconducting materials in DC and low-frequency domains, such as superconducting magnets and cables [1, 2], has yielded significant economic benefits and substantially advanced the development of related scientific disciplines.

Charged particle accelerators are indispensable tools in various fields including high-energy physics, atomic and molecular physics, life and materials science, nuclear physics, and radionuclide research. SRF technology stands as one of the fundamental technologies in modern particle accelerators, being selected as the preferred technical approach for ongoing and planned accelerator projects [3, 4, 5, 6]. The current SRF accelerator relies on resonant cavities made from niobium (Nb), typically cooled by immersion in 2K liquid helium. The 2K refrigeration system, with its complex structure and low cooling efficiency, results in high operational costs and poses maintenance challenges for Nb-cavity-based SRF accelerators. As a result, their application is primarily confined to research facilities, hindering their transition to industrial use. Utilizing new materials with higher transition temperatures and novel heat transfer mechanisms in SRF applications is advantageous for increasing operational temperatures and reducing complexity. It is not only of great significance for super-large accelerators [6, 7, 8], but also is the premise and key for SRF technology to get rid of the service limitation of liquid helium and find future applications in smaller scientific research platforms such as compact light sources [9], photo-neutron sources [10], and industrial applications such as wastewater treatment [11] and medical isotope production [12]. This constitutes an inevitable trend in the evolution of SRF technology, representing a common direction of effort in the SRF field.

Scientists have investigated a variety of materials with higher transition temperatures than Nb for SRF applications, including $Nb_3Sn$, $MgB_2$, NbN, NbTiN, and even YBCO [13]. Among these, $Nb_3Sn$ is currently the only new material that genuinely demonstrates the 2K performance of Nb cavities at 4.2K or even higher temperatures under extreme conditions such as strong RF electromagnetic fields, sensitivity to magnetic flux pinning, and extremely low losses. It is regarded as one of the pivotal technologies for next-generation SRF accelerators.

Currently, $Nb_3Sn$ films grown by the vapor diffusion method have the best RF performance. The tin vapor diffusion method for growing $Nb_3Sn$ materials was initially employed by Saur and Wurm in 1962 [14], while researchers at Siemens AG pioneered the application of $Nb_3Sn$ thin films grown by vapor diffusion method in SRF during the 1970s [15]. After about half a century of development, the understanding and optimization of the process for growing $Nb_3Sn$ thin films by tin vapor diffusion method gradually deepened, and the performance of the $Nb_3Sn$ thin film SRF cavity was continuously improved. Especially in 2013, Cornell University successfully coated a high-performance $Nb_3Sn$ thin film SRF cavity without Q-slope limitations using tin vapor diffusion method [16]. It quickly became a frontier hot spot in the SRF field and continues to this day. Subsequently, both FNAL [17] and JLab [18] also achieved successful coatings of $Nb_3Sn$ thin film SRF cavities with high RF performance using the tin vapor diffusion method.

Exploratory research in the practical applications of $Nb_3Sn$ thin film SRF cavities has primarily advanced in two directions. On the one hand, G. Eremeev carried out research of 1.5GHz 5 cell $Nb_3Sn$ thin film SRF cavity in engineering application under liquid helium immersion conditions. After continuous optimization of the assembly process, two 1.5 GHz 5-cell $Nb_3Sn$ thin film SRF cavities were assembled into a cavity pair and one of the cavities can basically reach the baseline level before assembly [19]. On the other hand, FNAL proposed the concept of conduction cooling of the $Nb_3Sn$ thin film SRF cavity by connecting the cavity outer surface to the commercial cryocooler with thermally conductive link in 2015 [20]. Subsequently, FNAL's 650MHz single-cell [19], Cornell University's 2.6GHz single-cell [21], and JLab's 952.6MHz single-cell $Nb_3Sn$ thin film SRF cavities [22] all attained a usable acceleration gradient of 10MV/m under conduction cooling conditions. Over the past two years, JLab, FNAL, and Cornell University have proposed several designs for high-current, high-power conduction cooling SRF accelerators in the energy range of 1-10MeV for different industrial applications [23]. In addition, KEK [24], PKU [25] and IHEP [26] also followed up the research of $Nb_3Sn$ thin film SRF cavity coating and application. However, whether based on liquid helium immersion cooling or LHe-free conduction cooling, the practical application of $Nb_3Sn$ thin film SRF cavities is still in the exploratory research phase. There is an urgent need for principal verification through accelerator system integration testing and stable beam acceleration.

IMP initiated the coating of $Nb_3Sn$ SRF cavities using the vapor diffusion method at the end of 2018. The application-oriented $Nb_3Sn$ coating apparatus capable of coating production-type multi-cell cavities has been built and achieved stable operation. Following studies into the mechanism of the observed non-uniform distribution of tin nuclei on the Nb surface with tin chloride ($SnCl_2$) [27], as well as an enhanced understanding of the coating process [28], the maximum accelerating gradient ($E_{acc,max}$) of the 1.3 GHz single cell $Nb_3Sn$ SRF cavity coated at IMP with optimized recipe reached 18.03 MV/m and the unloaded quality factor ($Q_0$) at $E_{acc}$ = 12 MV/m and 4.2 K exceeds 1E10. Since 2021, IMP has embarked on the study of LHe-free SRF technology, utilizing GM cryocoolers to drive the $Nb_3Sn$ cavity via conduction cooling. Under conduction cooling conditions, the $Q_0$ of the 1.3GHz single cell $Nb_3Sn$ thin film SRF cavity at low field reaches 9E9, and the $Q_0$ at $E_{acc,max}$ =6.6MV/m is higher than 4E9 [29]. For the engineering application of $Nb_3Sn$ thin film SRF cavities in the CiADS project, IMP designed and constructed a conduction-cooled $Nb_3Sn$ demonstration SRF accelerator to achieve principal verification of stable electron beam acceleration. The details will be described below.

**2. Design of the accelerator**

For the consideration of engineering application planning of $Nb_3Sn$ thin film SRF cavity in the CiADS project, a 650MHz 5 cell elliptical cavity was coated using vapor diffusion method for the acceleration of electron beam. Electron beam with macropulse length of 0.5-2.0μs and variable repetition rate is generated by a gridded thermionic cathode, followed by a cathode-anode potential of 60kV. The electron beam is focused by a solenoid and directed into the $Nb_3Sn$ cavity, where it undergoes simultaneous acceleration and longitudinal bunching. The electron beam current prior to entering the cavity is measured by ACCT. Following acceleration within the cavity and departure from it, the electron beam is once again focused by the downstream solenoid. On one hand, the electron beam is transmitted in a straight path, with the beam spot measured by a fluorescent target downstream of the solenoid, and ultimately collected by Dump1 at the terminal. On the other hand, the focused electron beam is deflected by 60 degrees using a dipole magnet, passing through a slit, and collected by Dump2 at another terminal to acquire the energy distribution of the accelerated electron beam. Schematic layout of the conduction cooled $Nb_3Sn$ demo SRF electron accelerator is shown in Figure 1.

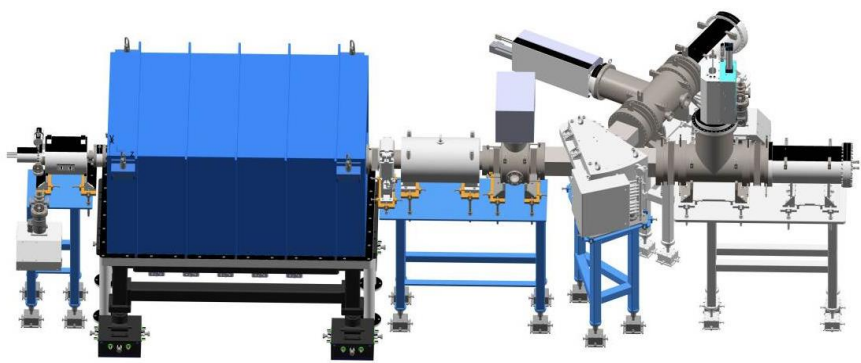

Fig. 1. Schematic layout of the conduction cooled $Nb_3Sn$ demo SRF electron accelerator

In the aforementioned design, there is no buncher preceding the entry of the DC electron beam into the cavity, and the β value of the 650MHz 5-cell $Nb_3Sn$ thin film SRF cavity is 0.82. Consequently, it is unavoidable that the electron beam will experience beam loss during the process of acceleration and transmission. To enhance transmission efficiency and minimize electron beam bombardment on the cavity's inner wall, the distance between the cryomodule's input port and the cavity flange is designed to be as short as possible. Additionally, the diameter of the cryomodule's output beampipe is relatively large. Simulation results indicate that at $E_{pk}$=13.78MV/m, the beam energy reaches 4.38MeV, with an energy dispersion of 6% and a transmission efficiency of 38.1%. Figure 2 depicts the envelope evolution and beam loss distribution during electron beam transmission, respectively.

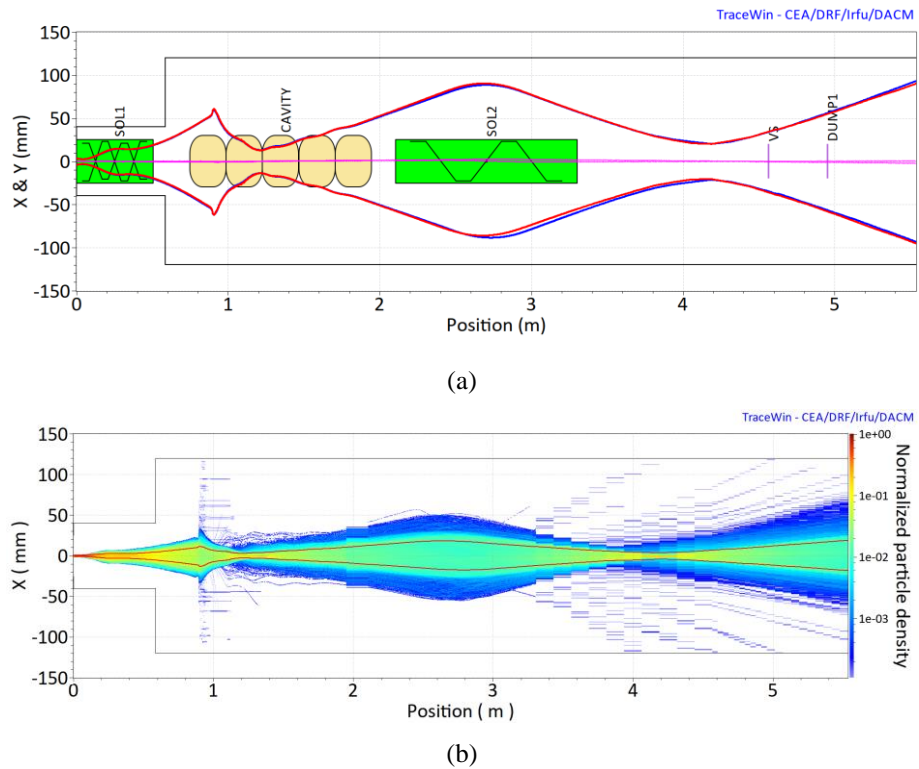

(a)

(b)

Fig. 2. Envelope evolution (a) and beam loss distribution (b) during electron beam transmission. The primary loss of the electron beam occurs within the beampipe between the cryomodule input port and the cavity flange, as well as within the upstream beamtube of the cavity.

**3. Construction of the accelerator**

Figure 3 (a) depicts the coating conditions of the 650MHz 5-cell $Nb_3Sn$ thin film SRF cavity in this paper, which refer to the previously optimized recipe at IMP. During the film growth stage, the temperature of the tin source heater is set to 1280 degrees, with a growth duration of 2 hours. Subsequently, the power supply to the tin source heater is switched off, and the cavity is maintained at 1100 degrees for 6 hours. Figure 3 (b) shows that the inner surface of the cavity is uniformly covered with typical gray $Nb_3Sn$ thin films to the naked eye. Following ultrasonic cleaning and high-pressure water rinsing (HPR), the cavity, coupler, vacuum pipeline, and valve linked to the cryomodule are integrated and assembled onto the auxiliary fixture bracket. Taking into account the stress-strain sensitivity of $Nb_3Sn$ film and the high vacuum environment surrounding the cavity in LHe-free conditions, we opted not to conduct vacuum leak detection on the integrated assembly system in the cleanroom. The integrated cavity system is depicted in Figure 4.

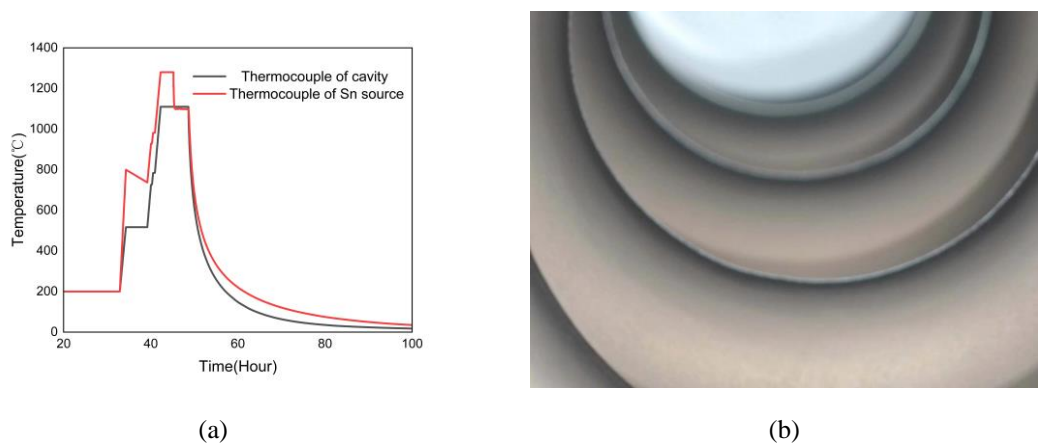

(a) (b)

Fig. 3. (a) Coating recipe of 650MHz 5 cell $Nb_3Sn$ thin film SRF cavity, (b) inner surface of the cavity after coating.

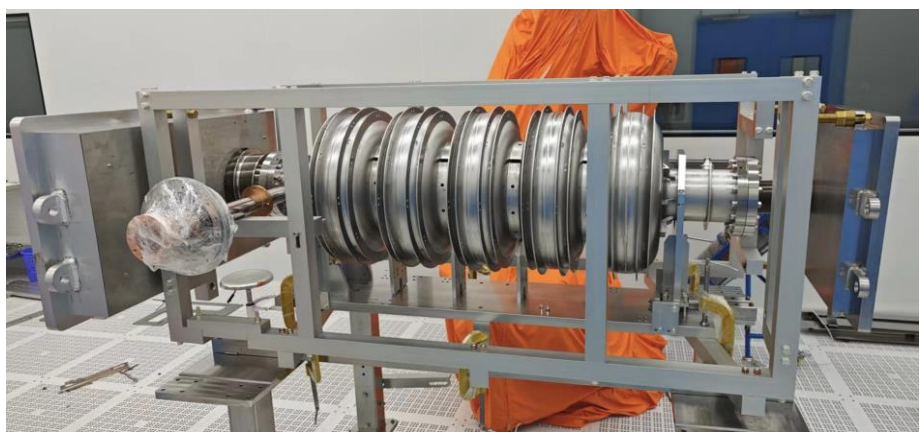

Fig. 4. Assembly of the cavity, coupler, vacuum pipeline, and valve connected to the cryomodule on the auxiliary fixture bracket.

The conduction cooling structure has been optimized and refined leveraging our past research experience, synergized with the current research status. Each cell of the cavity's outer surface is welded with four Nb rings, and each beamtube is also welded with one Nb ring in an effort to enhance cooling efficiency. To further enhance temperature distribution uniformity during the cooling process, six sets of holes for connecting with copper braided tapes are evenly distributed on each Nb ring, with the holes of adjacent rings staggered at a 30° angle difference. Meanwhile, the two ends of the copper braided tapes are directly connected to the second stage cold head of the cryocooler and the Nb ring to avoid excessive contact thermal resistance during transfer. Figure 5 shows the image after connecting the cavity Nb rings with the second stage cold head using copper braided tapes. The pumping ratio was set based on the volume between the cavity and the cryomodule. Then, the cavity and cryomodule are simultaneously vacuumed at a slow speed. The pressure difference between the inside and outside of the cavity is reduced as much as possible to avoid the cavity being squeezed and deformed during the vacuum process.

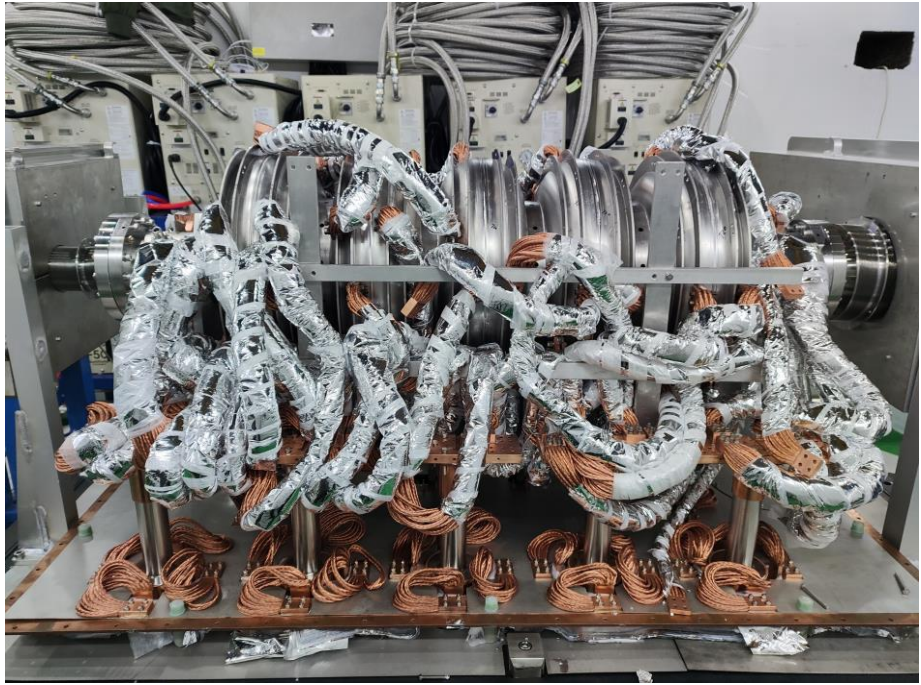

Fig. 5. The visual representation following the integration of the cavity's Nb rings with the second stage cold head using copper braided tapes.

## 4. Commissioning of the accelerator
### *4.1 Cooldown process*

Ten GM cryocoolers were used to cool the 650MHz 5 cell $Nb_3Sn$ thin film SRF cavity through conduction cooling. When the temperature drops to 25K, the second stage cold head of the cryocoolers remain at 25K for one hour to ensure adequate cooling of the cyromodule. Through high-precision cooperative control, 10 cryocoolers are simultaneously cooled from 25K at a rate of 10 minutes per Kelvin. The synchronization accuracy at the onset of cooling, as well as the temperature deviation range among the 10 cryocoolers during the slow cooling process across the sensitive temperature zone of 18K, are both below ±0.2K. As shown in Figure 6 (a), when the temperature of the cavity is ~17.89K and ~8.97K, the fluxgates located in the middle cell and the beam tube position exhibit distinct magnetic expulsion characteristics of both the $Nb_3Sn$ film and the underlying substrate of Nb metal. When the temperature of the second stage cold head is lower than 6K, the compensation control of the high precision heater is turned off, and the cavity begins to cool naturally. Figure 6 (b) shows the temperature distribution at different positions of the cavity and the second stage cold head of the cryocoolers under thermal equilibrium state. The input pipe diameter of the cryomodule is small, resulting in minimal radiation leakage heat. Conversely, the output pipe diameter is wide, leading to significant radiation leakage heat. Consequently, the temperature at downstream locations of the cavity is higher compared to upstream locations. Nonetheless, the temperature difference between different positions of the cavity and the adjacent second stage cold head typically remains below 0.3K. This indicates that our cooling structure has a good connection and the overall thermal resistance is small. In addition, according to the temperature distribution and cooling curve of the cryocooler, the static heat leakage of the cryomodule can be estimated to be less than 1.5W.

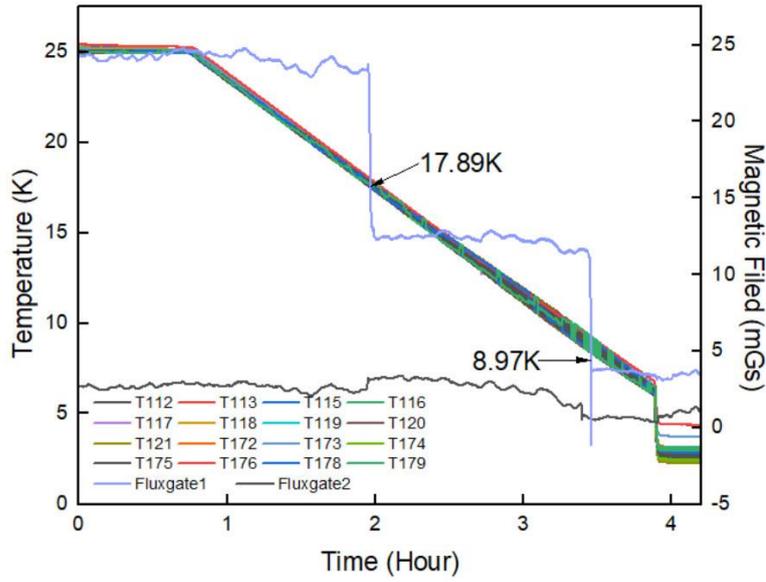

(a)

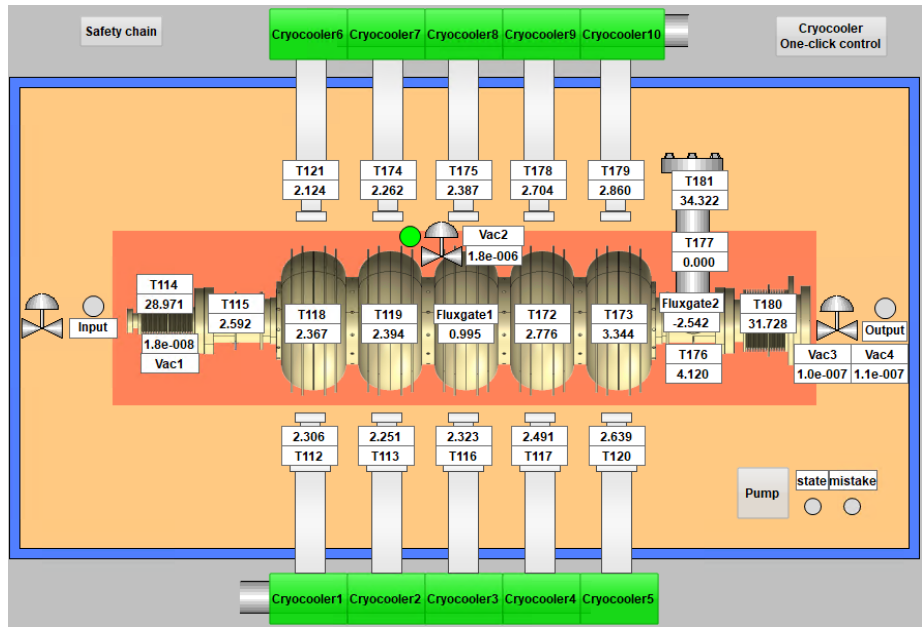

(b)

Fig. 6. (a) The cooling process and characteristic magnetic flux expulsion of the cryomodule, (b) temperature distribution of the cryomodule at thermal equilibrium state.

*4.2 Horizontal test*

After cooling down, horizontal tests were conducted on the LHe-free cryomodule in both pulse and continuous wave (CW) modes. A 650MHz 5kW solid state power source is used to feed power to the cavity via a coupler through a 3-1/8-inch coaxial transmission line. There is only one cavity in this accelerator. Therefore, we choose to operate the cavity in a self-excited mode in an open-loop state during the horizontal test. In pulse mode, with an input power pulse width of 20ms, a stable field can be established within the cavity. The duty cycle can be augmented by elevating the repetition rate of the incident pulses. Under duty cycles of 2%, 4%, 10%, 20%, and 40%, the cavity can stably operate at $E_{pk}$=14.84MV/m. However, in CW mode, the field at which the cavity can stably operate drops to $E_{pk}$=6.02MV/m. Table 1 lists the stable operation records spanning approximately 1 hour in both pulse mode and CW mode. In pulse mode, as the duty cycle increases, the quench field slightly decreases. The primary limiting factor is thermal breakdown caused by the failure to promptly dissipate the heat generated by field emission. In CW mode, although there is no field emission due to the lower RF field. The limiting factor remains thermal breakdown, stemming from the failure to promptly dissipate the heat generated by the RF field in CW mode. Therefore, for LHe-free SRF technology based on conduction cooling, enhancing heat transfer efficiency emerges as a crucial avenue for enhancing system thermal stability and operational gradient.

Table 1: Stable operation of the cryomodule in both pulse mode and CW mode

| Operation mode | $E_{pk}$ (MV/m) | Operating time (min) | $E_{pk}$@quench (MV/m) |
|---|---|---|---|
| 0.02s@1s 1Hz | 14.90 | 60 | 15.48 |
| 0.02s@0.5s 2Hz | 14.84 | 51 | 15.37 |
| 0.02s@0.2s 5Hz | 14.84 | 56 | 14.55 |
| 0.02s@0.1s 10Hz | 14.90 | 61 | 15.08 |
| 0.02s@0.05s 20Hz | 14.84 | 40 | 14.96 |
| CW | 6.02 | 60 | 6.25 |

Through high precision thermal compensation control, the second stage cold heads of the cryocoolers can be maintained at 4.2K. When RF power is supplied to the cavity, electromagnetic fields will induce heat losses on the cavity surface. To uphold the cold heads at 4.2K, their thermal compensation will be adjusted accordingly. By gauging the variance in compensation heat, the heat load of the cavity at a specific gradient can be computed. Figure 7 depicts measurements of thermal compensation for the cold head under various acceleration gradients. Table 2 lists the calculated heat load of cavity operated in specific acceleration gradients.

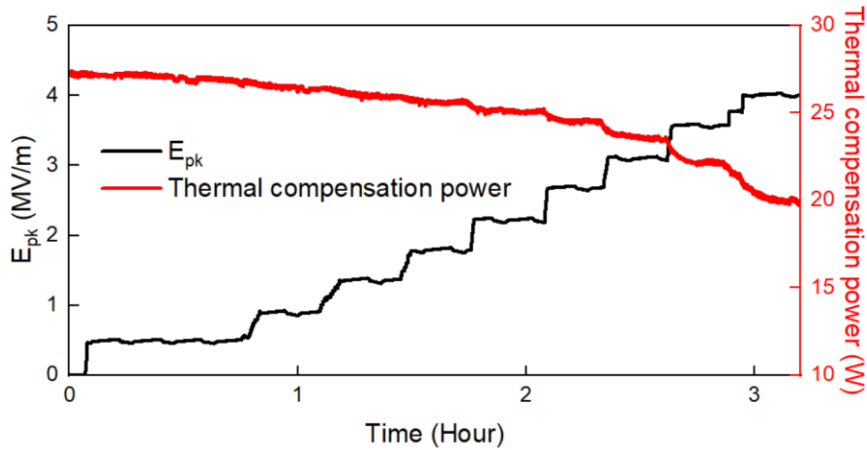

Fig. 7. Changes of thermal compensation of the cold head with the acceleration gradient

Table 2: Heat load of the cavity at different acceleration gradients

| $E_{pk}$ (MV/m) | Heat load (W) |
|---|---|
| 0.84 | 0.19 |
| 1.37 | 0.77 |
| 1.84 | 1.23 |
| 2.18 | 1.59 |
| 2.47 | 2.10 |
| 2.74 | 2.67 |
| 3.00 | 3.56 |
| 3.31 | 4.94 |
| 3.63 | 7.07 |

*4.3 Electron beam acceleration*

Beam acceleration experiments were conducted in both pulse mode and CW mode. In the subsequent experiments, the injected electron beams were all DC macropulses with an energy of 60kV and a pulse width of 2μs. Under the conditions of an input power pulse width of 20 milliseconds, a repetition frequency of 1 Hertz, a duty cycle of 2%, and $E_{pk}$=13.78MV/m, the average beam current within the macropulse detected by ACCT is 256mA. Figure 8 illustrates the beam loading effect as the electron beam passes through the cavity and the resultant beam spot formed upon bombardment of the fluorescence target after acceleration.

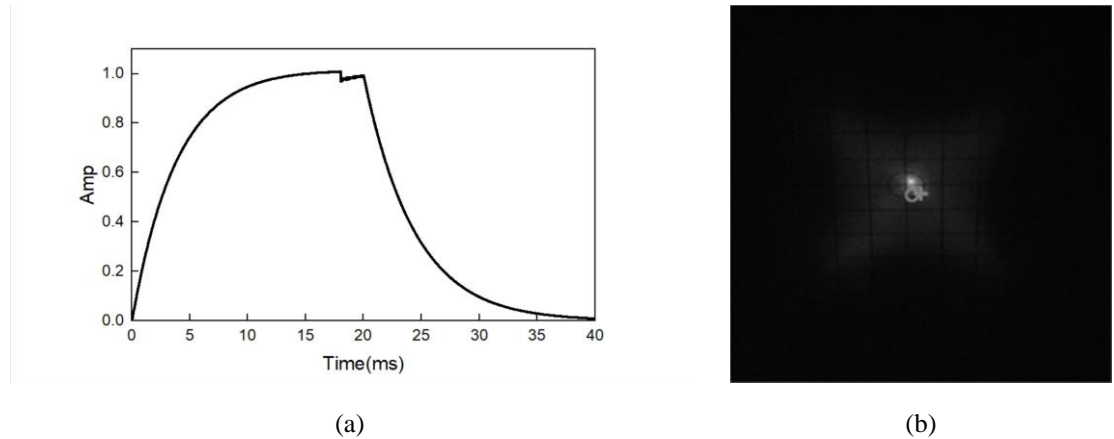

(a)          (b)

Fig. 8. (a) Beam loading effect when the electron beam pass the cavity, (b) beam spot formed by electron beam bombardment on fluorescent targets.

The average beam current within the macropulse collected by Dump1 is 113mA, with a transmission efficiency of 44.14%. By scanning the current of the bipolar magnet and recording the beam current collected by Dump2, the central energy of the electron beam is determined to be 4.25MeV, as depicted in Figure 9 (a). Subsequently, a slit is positioned in the center of the pipe after the deflection magnet, and the current of the bipolar magnet is scanned again to measure the fine energy structure near the central energy. The specific results are illustrated in Figure 9 (b). The results indicate that the half-height width of the fine energy structure is about 0.21MeV, with a corresponding energy dispersion of about 5.04%. The beam spot size, transmission efficiency, central energy, and energy dispersion agree well with the simulation results. Maintaining the injection parameters and RF duty cycle unchanged, the $E_{pk}$ value of the cavity is then increased to 14.90MV/m, resulting in an accelerated electron beam energy reaching 4.6MeV.

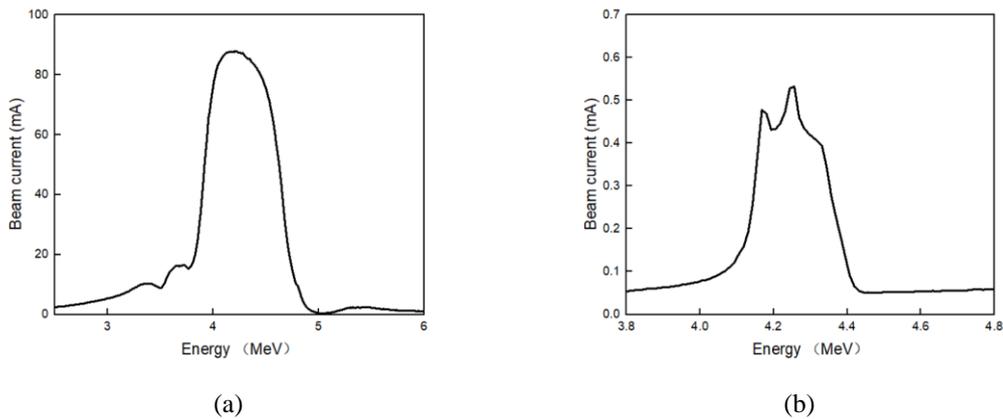

(a)　　　　　　　　　　　　　　　　(b)

Fig. 9. The central energy and fine energy structure obtained by scanning the current of the bipolar magnet.

Considering the heat load caused by beam loss and the emission capacity of the electron gun, the average current of the injected electron beams within the macropulse is reduced to 134mA, while the repetition frequency is increased to 20Hz. T By increasing the repetition rate of the input power pulses to elevate the duty cycle to 40%, and reducing the $E_{pk}$ value to 8.18MV/m, high-current beam acceleration tests in pulse mode were subsequently conducted. The average current of the electron beam collected by Dump1 reaches 70mA within the macropulse. Additionally, as depicted in Figure 10, under the aforementioned conditions, the acceleration and transmission of the electron beam remain stable over a time span of 40 minutes.

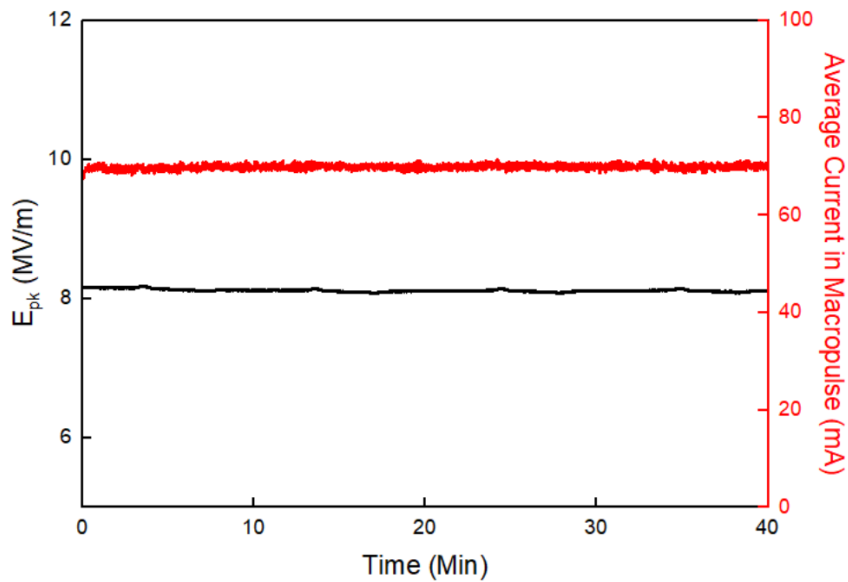

Fig. 10. Stable acceleration records of the electron beam with an average current of 70mA within the macropulse, operating at a repetition rate of 20Hz and an $E_{pk}$ of 8.18MV/m under a 40% duty cycle condition.

An important advantage of SRF accelerators is their capability to operate in CW mode. Therefore, we proceeded to conduct electron beam acceleration experiments in CW mode at $E_{pk}$=3.91MV/m. When injecting an electron beam with a repetition rate of 1Hz and an average beam current within the macropulse of 511mA, the average current collected by Dump1 within the macropulse reaches 235mA, with a transmission efficiency of 46.0%. The central energy, determined by scanning the current of the bipolar magnet, is approximately 1.57MeV. Figure 11 depicts the beam spot on the fluorescent target and the energy structure of the electron beam under the aforementioned conditions.

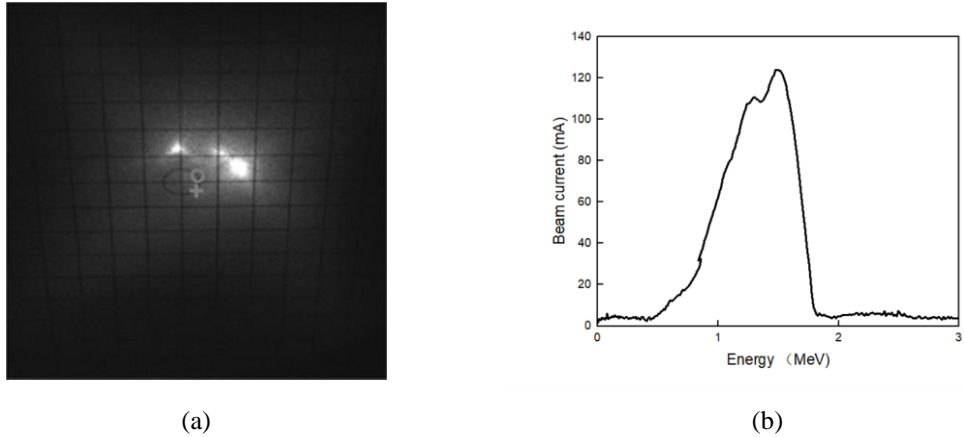

(a)          (b)

Fig.11. The beam spot on the fluorescent target and the energy structure of the electron beam in CW mode with an injected electron beam repetition rate of 1Hz.

Maintaining the $E_{pk}$ constant in CW mode, the repetition rate of the injected electron beam is increased to further evaluate the stability of high-current acceleration in CW mode. When the injected electron beam has a repetition rate of 50Hz and an average beam current of 230mA within the macropulse, the electron beam current collected by Dump1 after acceleration is ~120mA. Due to the limited emission capability of the electron gun under high repetition rate conditions, during a 10 minute stable testing process, the electron beam current collected by Dump1 decreased from 123mA to 117mA. Subsequently, the injected electron beam current is reduced to approximately 133mA, and their repetition rate is increased to 100Hz, 150Hz, and 200Hz respectively. Throughout each 10 minute test period, the electron beam current collected by Dump1 consistently remained stable at approximately 64mA. Figure 12 shows the stable acceleration record of the high repetition rate injected electron beam in CW mode.

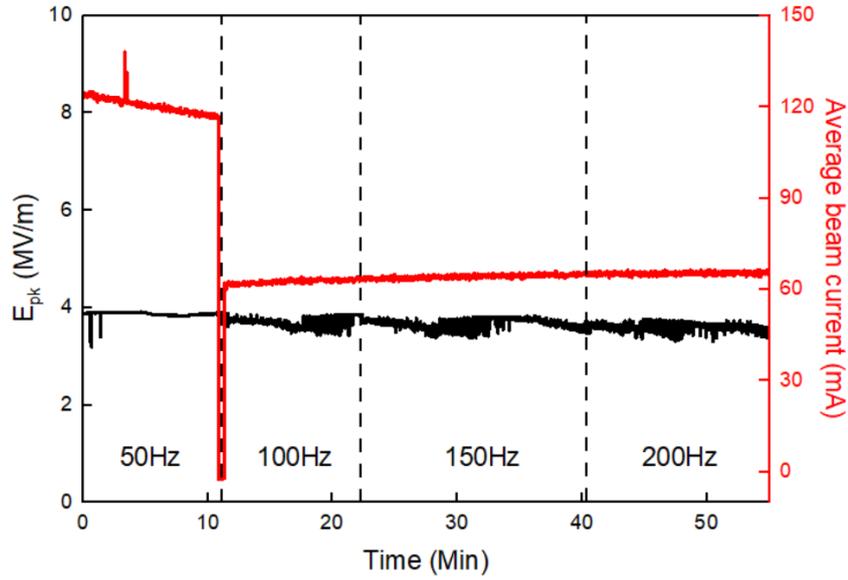

Fig. 12. Stable acceleration records in CW mode with $E_{pk}$=3.91MV/m, average beam current within macropulse of 64mA, and injected electron beam repetition rates of 50Hz, 100Hz, 150Hz, and 200Hz, respectively.

**5. Conclusion:**

This paper reports on the design, construction, and commissioning progress of a conduction-cooled $Nb_3Sn$ demonstration SRF accelerator at IMP:

1. A 650MHz 5 cell $Nb_3Sn$ thin film SRF cavity for electron beam acceleration was coated using tin vapor diffusion method. Ten GM cryocoolers were used to cool the cavity through conduction cooling. Efforts were made to address the reduction of connection thermal resistance and deformation stress during the integrated assembly of the LHe-free cryomodule.
2. High-precision and controllable cooling of the cavity has been achieved through collaborative local thermal compensation of the second stage cold heads. Horizontal tests were conducted in both CW and pulse modes with varying duty cycles, achieving stable operation at the hourly level at gradients of $E_{pk}$=6.02MV/m and 14.84MV/m, respectively.
3. In both pulse and CW modes, stable acceleration of the electron beam with high repetition rates and macropulse average currents of more than 60mA can be achieved. In the pulse mode, the average current of accelerated beam in the macropulse exceeds 100mA, and the central energy reaches 4.6MeV. In CW mode, the average current of accelerated beam in the macropulse exceeds 200mA, and the central energy reaches 1.57MeV.

The work of this article provides principal verification for the application of $Nb_3Sn$ thin film SRF cavities in both large-scale scientific facilities and and compact industrial accelerators.

**Acknowledgments**

This work is supported by the National Natural Science Foundation of China (No.12175283), Youth Innovation Promotion Association of Chinese Academy of Sciences (2020410) and Advanced Energy Science and Technology Guangdong Laboratory (HND20TDSPCD, HND22PTDZCD).